# Formulation of spinors in terms of gauge fields

S. R. Vatsya

648 Inverness Ave., London, ON, Canada, N6H 5R4
e-mail: raj.vatsya@gmail.com

## Abstract

It is shown in the present paper that the transformation relating a parallel transported vector in a Weyl space to the original one is the product of a multiplicative gauge transformation and a proper orthochronous Lorentz transformation. Such a Lorentz transformation admits a spinor representation, which is obtained and used to deduce the transportation properties of a Weyl spinor, which are then expressed in terms of a composite gauge group defined as the product of a multiplicative gauge group and the spinor group. These properties render a spinor amenable to its treatment as a particle coupled to a multidimensional gauge field in the framework of the Kaluza-Klein formulation extended to multidimensional gauge fields. In this framework, a fiber bundle is constructed with a horizontal, base space and a vertical, gauge space, which is a Lie group manifold, termed its structure group. For the present, the base is the Minkowski spacetime and the vertical space is the composite gauge group mentioned above. The fiber bundle is equipped with a Riemannian structure, which is used to obtain the classical description of motion of a spinor. In its classical picture, a Weyl spinor is found to behave as a spinning charged particle in translational motion. The corresponding quantum description is deduced from the Klein-Gordon equation in the Riemann spaces obtained by the methods of path-integration. This equation in the present fiber bundle reduces to the equation for a spinor in the Weyl geometry, which is close to but differs somewhat from the squared Dirac equation.

# I. Introduction

In the Riemannian spaces, a vector undergoing parallel displacement is required to retain its original length. Weyl's geometry is introduced in a manifold by abandoning this requirement of length invariance. Instead, it is assumed that the length $l$ of a vector $\ell$ at $x$ undergoes a change of $\delta l = \phi_\mu dx^\mu l$ under the parallel transport to a neighboring point $(x+dx)$, where the gauge potentials $\phi_\mu$ are some functions. This renders the underlying manifold into an affinely connected space with its affine connections determined by the displacement property of length and by requiring the connection coefficients to be symmetric. The affine connection coefficients determine the transport properties of the vectors and higher rank tensors but the transport properties of spinors are not so straightforward to obtain. However, guided by the transport properties of vectors, the transport properties of spinors in Weyl's geometry paralleling those of the vectors have also been developed [1].

It is shown in the present article that in the Weyl space with base being the Minkowski manifold $\mathcal{M}_4$, a vector displaced to a neighboring point is transformed into another one with the transformation being the product of a multiplicative gauge transformation and a proper orthochronous Lorentz transformation. The spinor representation of such a Lorentz transformation can be determined explicitly [2]. The representation for the present case is obtained and used to deduce the transportation properties of spinors undergoing parallel displacement in the Weyl geometry resulting in essentially the same as obtained earlier by different arguments [1]. The transformation group for the spinors is expressible as the direct product of a multiplicative length gauge group and a 6-dimensional non-commutative spinor group constituting the 7-dimensional gauge group together with its seven gauge potentials. These properties render the spinors amenable to their treatment as the scalar particles coupled to multidimensional gauge fields. This construction and conclusion are extendible to the vectors and higher rank tensors in the Weyl spaces with their corresponding structure groups and gauge potentials.

Motion of the particles coupled with general gauge fields has been widely studied in the framework of fiber bundles with the horizontal base space in general being a Riemannian space and the vertical space being the gauge group manifold constituting the structure group of the bundle [3-5]. The fiber bundle itself is equipped with a Riemannian structure constituting an underlying space for the classical description of motion as its geodesics [4,5]. For the present case, where the base is the Minkowski spacetime manifold and the structure group is the gauge group described above, the fiber bundle is still a curved Riemannian space. An equation describing the classical motion in this space is obtained here by a more straightforward method than the ones used in literature. Present procedure to study the motion of particles coupled to multidimensional gauge fields is extendible also to the case of a general Riemannian space being the horizontal space.

Since the Dirac equation resulted directly from a treatment of the relativistic quantum mechanical electron, its classical analogue is not readily available. While a classical description of the spinors is interesting in itself, it is also required for a calculation of the action, which is needed to deduce their quantum mechanical description by the path-integral method [6,7]. There has been considerable interest in formulating the Dirac equation in the framework of path-integrals from the beginning with Feynman introducing the zigzag trajectories [8]. Out of a variety of approaches and formulations, we mention a few relevant ones.

A characteristic of the Dirac equation, zitterbewegung shows up in the classical description based on a Lagrangian postulated by Barut and Zanghi [9], which Barut and Duru used to deduce the Dirac propagator [10]. Schulman formulated the spinors in terms of the path-integrals in a curved space [11]. In a recent formulation the mass term in the Lagrangian of a relativistic spinning top was modified by including the curvature scalar of the same Weyl space as the one arising here. The resulting Hamilton-Jacobi equation was then shown to be equivalent to the squared Dirac equation operating on a specific type of wavefunctions [12]. Intention for the present was not to obtain a classical description of the Dirac spinor. Instead, we deduce the spinors from the Weyl vectors exploiting their parallel transport properties. This naturally lends the Weyl spinors amenable to their treatment as the particles coupled to multidimensional gauge fields. This paves way for their classical description as particles in the Riemannian spaces and then their quantum treatment in the framework of path-integrals in curved spaces [11,13-15] yielding the Klein-Gordon equation in general Riemannian spaces. The Klein-Gordon equation for a Weyl spinor, i.e., in the present fiber bundle, reduces to an equation close to but differing somewhat from the squared Dirac equation.

In Sec. II, we introduce the Kaluza-Klein formulation for multidimensional gauge fields. The parallel displacement properties of spinors are obtained from those of the Weyl vectors in Sec. III. The treatment of Sec. IV shows that the transport properties of spinors can be expressed in terms of a gauge group and thus, they can naturally be considered particles in gauge fields. In Sec. V and Sec. VI, the classical and quantum mechanical descriptions of the Weyl spinors are presented, followed up with some concluding remarks in Sec. VII.

## II. Gauge fields

Original Kaluza-Klein construction for a charged particle coupled to an electromagnetic field has been extended to particles coupled to multidimensional gauge fields in the framework of fiber bundles [3-5]. In this approach, the translational motion is described in a base space, which is usually a Riemannian manifold with its metric $\left[g_{\mu\nu}\right]$; and the internal motion, in a vertical, gauge space, which is a Lie group manifold, called its structure group, endowed with a metric $\left[g_{\alpha'\beta'}\right]$. The bundle is constructed by attaching the vertical space to the base. The composite space is then endowed with a Riemannian structure by defining a metric $\left[\hat{g}_{ab}\right]$ in terms of $[g_{\mu\nu}]$ and $\left[g_{\alpha'\beta'}\right]$. The indices $\mu, \nu, \lambda, \eta, ...$ with values $0,1,2,3$; and $\alpha', \beta', \gamma', \delta', ...$ with values $4, 5, ..., N_g + 3$; and $a, b, c, d, ....$ with values $0, 1, ..., N_g + 3$ are used here to label the coordinates on the base; the group manifold of dimension $N_g$, and the composite fiber bundle; respectively.

Let $\{e_{\alpha'}\}$ denote a basis in the Lie algebra of the structure group with its structure constants $C_{\alpha'\beta'}{}^{\gamma'}$, defined by $[e_{\alpha'}, e_{\beta'}] = C_{\alpha'\beta'}{}^{\gamma'} e_{\gamma'}$, and a metric $[g_{\alpha'\beta'}] = [e_{\alpha'} \bullet e_{\beta'}]$ with the thick dot denoting the scalar product on the algebra. Weyl's parallel displacement rule $\delta l = \phi_\mu dx^\mu l$ admits a natural extension to the multidimensional gauge fields, which is c, where the gauge potentials $\varphi_\mu^{\alpha'}$ are some functions on the group manifold as well as on the base. In the matrix representations of $e_{\alpha'}$, the length $l$ is represented by a column vector. Connections on the bundle are given by the Lie algebra valued 1-forms $dx^\mu \varphi_\mu^{\alpha'} e_{\alpha'}$, which connect the base and the gauge space. In the extended Kaluza-Klein formulation, the metric $[\hat{g}_{ab}]$ and its inverse $[\hat{g}^{ab}]$ on the bundle are given by [4,5]

$$[\hat{g}_{ab}] = \begin{bmatrix} [g_{\mu\nu} + g_{\alpha'\beta'} \varphi_\mu^{\alpha'} \varphi_\nu^{\beta'}] & [g_{\alpha'\beta'} \varphi_\mu^{\alpha'}] \\ [g_{\alpha'\beta'} \varphi_\nu^{\beta'}] & [g_{\alpha'\beta'}] \end{bmatrix};$$

$$[\hat{g}^{ab}] = \begin{bmatrix} [g^{\mu\nu}] & -[g^{\mu\nu} \varphi_\mu^{\alpha'}] \\ -[g^{\mu\nu} \varphi_\nu^{\beta'}] & [g^{\alpha'\beta'} + g^{\mu\nu} \varphi_\mu^{\alpha'} \varphi_\nu^{\beta'}] \end{bmatrix}. \quad (1)$$

Infinitesimal arclength $d\tau$ defined by $d\tau^2 = dx_a dx^a$ is calculated from the metric, yielding

$$d\tau^2 = g_{\mu\nu} dx^\mu dx^\nu + g_{\alpha'\beta'} (dx^{\alpha'} + \varphi_\mu^{\alpha'} dx^\mu)(dx^{\beta'} + \varphi_\nu^{\beta'} dx^\nu). \quad (2)$$

### III. Spinors in Weyl geometry

We take the Minkowski spacetime $\mathcal{M}_4$ for the base with its diagonal metric defined by

$$g_{00} = 1, \ g_{\mu\mu} = -1, \ \mu = 1, 2, 3; \ g_{\mu\nu} = 0, \ \mu \neq \nu.$$

Weyl's geometry is introduced in it by requiring that $\delta l = \phi_\mu dx^\mu l$, yielding an affinely connected space with connections $\Gamma^\lambda_{\mu\nu}$ prescribing the transport property for vectors by $\delta \ell^\lambda = -\Gamma^\lambda_{\mu\nu} dx^\nu \ell^\mu$. Further requirement of symmetry $\Gamma^\lambda_{\mu\nu} = \Gamma^\lambda_{\nu\mu}$ determines the affine connection coefficients uniquely [16]:

$$\Gamma^\lambda_{\mu\nu} = \Gamma^\lambda_{\nu\mu} = -\left[\delta^\lambda_\mu \phi_\nu + \delta^\lambda_\nu \phi_\mu - g_{\mu\nu}\phi^\lambda\right]; \tag{3}$$

where $\delta^\lambda_\mu = g^{\lambda\nu} g_{\nu\mu}$ is the Kronecker delta. It should be mentioned that the symmetry of affine connections is not necessary to maintain Weyl's length gauge condition and the following treatment can be adjusted to accommodate nonsymmetrical affine connections.

It follows from (3) that under the displacement $x \rightarrow (x + dx)$, a vector $\ell$ transforms into $\ell'$ given by

$$\ell'^\lambda = \left[\delta^\lambda_\mu + dx^\nu \left(\delta^\lambda_\mu \phi_\nu + \delta^\lambda_\nu \phi_\mu - g_{\mu\nu}\phi^\lambda\right)\right]\ell^\mu;$$

i.e.,

$$\ell' = \left[1 + dx^\nu \phi_\nu + dx^\nu A_\nu\right]\ell;$$

where $A_\nu$ is a matrix with elements $(A_\nu)^\lambda_\mu = \left(\delta^\lambda_\nu \phi_\mu - g_{\mu\nu}\phi^\lambda\right)$. Since

$$\left[1 + dx^\nu \phi_\nu + dx^\nu A_\nu\right] = \left(1 + dx^\nu \phi_\nu\right)\left(1 + dx^\nu A_\nu\right), \tag{4}$$

up to the first order in $dx$, the group of transformations generated by the Lie algebra element $\left(dx^\nu \phi_\nu + dx^\nu A_\nu\right)$ is the direct product of the groups generated by $\phi_\nu dx^\nu$ and $A_\nu dx^\nu$. Treatment on the groups will be restricted to elements close to the identity, i.e., the equalities are valid up to the first order in $dx$, which is sufficient. All group elements can be generated by multiplying such elements.

The group generated by $\phi_\nu dx^\nu$ is the familiar multiplicative gauge group having an effect of changing the magnitude of each component of $\ell$. It can be seen by straightforward algebra that the group generated by $A_\nu dx^\nu$ preserves the length, i.e., with $\ell' = \left[1 + dx^\nu A_\nu\right]\ell$, we have $\ell'_\lambda \ell'^\lambda = \ell_\lambda \ell^\lambda$. Thus, this is a group of the Lorentz transformations. It can be checked that $\left(1 + dx^\nu A_\nu\right)^{00} = 1 > 0$, i.e., it is orthochronous; and the determinant of $(1+dx^\nu A_\nu)=1$, implying that the transformation is also proper. Further insight into this group can be gained by expressing

$$\left[1 + dx^\nu A_\nu\right]_{\lambda\mu} = g_{\lambda\eta}\left[1 + dx^\nu A_\nu\right]^\eta_\mu$$

as

$$\begin{aligned}\left[1 + dx^\nu A_\nu\right]_{\lambda\mu} &= \left[g_{\lambda\mu} + dx^\nu \phi^\eta \left(g_{\nu\lambda}g_{\mu\eta} - g_{\mu\nu}g_{\lambda\eta}\right)\right] \\ &= \left(g_{\lambda\mu} - dx^\nu \phi^\eta Z_{\lambda\mu(\nu\eta)}\right)\end{aligned} \quad (5)$$

The matrix $\left[g_{\lambda\mu} - \varepsilon'^{\nu\eta}Z_{\lambda\mu(\nu\eta)}\right]$ corresponds to a rotation through an angle of $\varepsilon'^{\nu\eta}$ in $x^\nu x^\eta$ plane [17]. The angle of rotation in (5) is equal to $dx^\nu \phi^\eta$. Thus, $A_\nu dx^\nu$ generates a group of the Lorentz rotations. Spinor representations of the proper, orthochronous Lorentz groups are available in literature [2,17,18]. We reproduce the details needed for clarity and obtain the spinor representation of the present group.

There is one to one correspondence between the vectors $\ell$ in $\mathcal{M}_4$ and the spinors $\mathcal{X}$ of rank 2, defined by $\mathcal{X} = \ell_\mu \sigma^\mu$, where $\sigma^0$ is the 2-dimensional identity matrix and $\sigma^\alpha$, $\alpha = 1,2,3$, are the Pauli matrices with the commutation properties

$$\left[\sigma_\alpha, \sigma_\beta\right] = 2i\varepsilon_{\alpha\beta}{}^\gamma \sigma_\gamma, \quad \alpha,\beta,\gamma = 1,2,3; \quad (6)$$

where $\varepsilon_{\alpha\beta\eta}$ is the Levi-Civita tensor density in 3-space, normalized so that $\varepsilon^{123}=1$. For a real vector $\ell$, $\mathcal{X}$ is clearly selfadjoint. Let $\bar\sigma^0 = \sigma^0$ and $\bar\sigma^\alpha = -\sigma^\alpha$, $\alpha = 1,2,3$; i.e., the 4-vectors $\sigma$ and $\bar\sigma$ are obtained from each other by spatial reflections. The vector $\ell$ is recovered from $\mathcal{X}$ by $\ell^\mu = Tr(\bar\sigma^\mu \mathcal{X})/2$, where $Tr$ denotes the trace of a matrix. It follows by direct calculations that

$$\sigma^\mu \bar\sigma^\nu + \sigma^\nu \bar\sigma^\mu = \bar\sigma^\mu \sigma^\nu + \bar\sigma^\nu \sigma^\mu = 2g^{\mu\nu}. \quad (7)$$

Since

$$g^{\mu\nu}\ell_\nu = \ell^\mu = \frac{1}{2}Tr(\bar\sigma^\mu \mathcal{X}) = \frac{1}{2}Tr(\bar\sigma^\mu \ell_\nu \sigma^\nu) = \frac{1}{2}Tr(\bar\sigma^\mu \sigma^\nu)\ell_\nu,$$

for an arbitrary vector $\ell$, we have that $g^{\mu\nu} = Tr(\bar\sigma^\mu \sigma^\nu)/2$, providing an alternative expression for $g^{\mu\nu}$.

For a given proper, orthochronous Lorentz transformation $\Omega$, one can find $\Lambda$ such that [2,17]

$$\Omega^{\mu}{}_{\nu}\, \sigma_{\mu} = \Lambda\, \sigma_{\nu}\, \Lambda^{\dagger}; \quad \Omega_{\nu}{}^{\mu}\, \bar{\sigma}_{\mu} = \Lambda^{\dagger}\, \bar{\sigma}_{\nu}\, \Lambda; \tag{8}$$

where the dagger denotes the adjoint. The transformation $\Lambda$ is given explicitly by [2]

$$\Lambda = \frac{\Omega_{\mu\nu}\, \sigma^{\mu}\, \bar{\sigma}^{\nu}}{\left[\Omega_{\mu\nu}\, \Omega_{\gamma\eta}\, \sigma^{\mu}\, \bar{\sigma}^{\nu}\, \sigma^{\eta}\, \bar{\sigma}^{\gamma}\right]^{1/2}}. \tag{9}$$

This yields a two to one homomorphism: $\pm\Lambda \leftrightarrow \Omega(\Lambda)$, providing a spinor representation of the Lorentz group. For the element $(1 + dx^{\nu} A_{\nu})$, we have

$$\Lambda = \left[1 + \frac{1}{4}\, dx^{\mu}\phi^{\nu}\, \left(\sigma_{\mu}\bar{\sigma}_{\nu} - \sigma_{\nu}\bar{\sigma}_{\mu}\right)\right]. \tag{10}$$

Now, consider the spinor $\mathcal{X} = \ell_{\mu}\sigma^{\mu}$ corresponding to a null vector $\ell$, defined by $\ell_{\mu}\ell^{\mu} = 0$. The spinor $\mathcal{X}$ is expressible as the product of a spinor $\theta$ of first rank and its adjoint as

$$\mathcal{X} = \theta\, \theta^{\dagger} = 2\begin{pmatrix}\xi \\ \eta\end{pmatrix}\begin{pmatrix}\xi^{*} & \eta^{*}\end{pmatrix},$$

where the star denotes the complex conjugate. The vector $\ell$ can be recovered from $\theta$ from $\ell^{\mu} = \theta^{\dagger}\bar{\sigma}^{\mu}\theta$ [18]. Let $\ell'$ be a null vector obtained from $\ell$ under a Lorentz transformation $\Omega$, i.e.,

$$(\Omega\, \ell)^{\mu} = \Omega^{\mu}{}_{\nu}\, \ell^{\nu} = \ell'^{\mu}.$$

It follows from (8) that

$$\begin{aligned}\ell'^{\mu} &= \Omega^{\mu}{}_{\nu}\, \ell^{\nu} = \Omega^{\mu}{}_{\nu}\, \theta^{\dagger}\, \bar{\sigma}^{\nu}\, \theta = \theta^{\dagger}\, \Lambda^{\dagger}\, \bar{\sigma}^{\mu}\, \Lambda\, \theta \\ &= (\Lambda\, \theta)^{\dagger}\, \bar{\sigma}^{\mu}\, (\Lambda\, \theta) = \theta'^{\dagger}\, \bar{\sigma}^{\mu}\, \theta';\end{aligned} \tag{11}$$

i.e., $\Omega$ transforms the spinor $\theta$ into $\theta' = (\Lambda\theta)$. Action of the multiplicative element $\vartheta$ can be incorporated into (11) by noting that

$$\ell''^{\mu} = \vartheta\, \ell'^{\mu} = \left(\theta'\, \sqrt{\vartheta}\right)^{\dagger} \bar{\sigma}^{\mu} \left(\sqrt{\vartheta}\, \theta'\right) = \vartheta''^{\dagger}\, \bar{\sigma}^{\mu}\, \vartheta'' \qquad (12)$$

In general, the dual spinor of $\vartheta''$ may not be equal to $\vartheta''^{\dagger}$. In such cases, (12) can be adjusted to obtain $\ell''^{\mu} = \vartheta''^{d} \bar{\sigma}^{\mu} \vartheta''$ with an appropriate dual $\vartheta''^{d}$ of $\vartheta''$ [1].

It follows from (10) and (12) that under the transportation from $x \to (x+dx)$ a 2-component spinor $\psi_x^2 \to \psi_{x+dx}^2$, where

$$\psi_{x+dx}^2 = \left[1 + \frac{1}{2}\, dx^{\mu} \phi_{\mu}\, \sigma_0 + \frac{1}{4}\, dx^{\mu} \phi^{\nu} \left(\sigma_{\mu}\bar{\sigma}_{\nu} - \sigma_{\nu}\bar{\sigma}_{\mu}\right)\right] \psi_x^2. \qquad (13)$$

The transformation of (13) can be adjusted for the reflection in space by interchanging $\bar{\sigma}$ and $\sigma$ as they are the space reflections of each other, yielding the corresponding transformation property for the space-reflected 2-component spinor $\hat{\psi}^2$, given by

$$\hat{\psi}_{x+dx}^2 = \left[1 + \frac{1}{2}\, dx^{\mu} \phi_{\mu}\, \sigma_0 + \frac{1}{4}\, dx^{\mu} \phi^{\nu} \left(\bar{\sigma}_{\mu}\sigma_{\nu} - \bar{\sigma}_{\nu}\sigma_{\mu}\right)\right] \hat{\psi}_x^2. \qquad (14)$$

With the Dirac $\gamma$ - matrices defined by

$$\gamma^{\mu} = \begin{bmatrix} 0 & \sigma^{\mu} \\ \bar{\sigma}^{\mu} & 0 \end{bmatrix},$$

it follows by direct computation that

$$[\gamma_{\mu}, \gamma_{\nu}] = \begin{bmatrix} (\sigma_{\mu}\bar{\sigma}_{\nu} - \sigma_{\nu}\bar{\sigma}_{\mu}) & 0 \\ 0 & (\bar{\sigma}_{\mu}\sigma_{\nu} - \bar{\sigma}_{\nu}\sigma_{\mu}) \end{bmatrix} = -4\, i\, s_{\mu\nu}, \qquad (15)$$

From (13), (14) and (15), the transportation properties of 4-component spinors can be expressed as

$$\psi_x^4 = \begin{bmatrix} \psi_x^2 \\ \hat{\psi}_x^2 \end{bmatrix} \to \psi_{x+dx}^4 = \left\{1 + \frac{1}{2}\, dx^{\mu}\, \phi_{\mu}\, e_0 - i\, dx^{\mu}\, s_{\mu\nu}\phi^{\nu}\right\} \psi_x^4$$
$$= \left\{1 + \frac{1}{2}\, dx^{\mu}\, \gamma_{\mu}\gamma_{\nu}\, \phi^{\nu}\right\} \psi_x^4 \qquad ; \quad (16)$$

where $e_0$ is the identity element.

Adler [1] obtained a similar transport property

$$\psi_x^4 = \begin{bmatrix} \psi_x^2 \\ \hat{\psi}_x^2 \end{bmatrix} \rightarrow \psi_{x+dx}^4 = \{1 + dx^\mu \zeta_\mu e_0 - i\, dx^\mu s_{\mu\nu} \phi^\nu\} \psi_x^4; \quad (17)$$

by different considerations including some intuitive arguments and analogies with vectors, where the functions $\zeta_\mu$ are arbitrary to a large extent. To assign a definite value to $\zeta_\mu$, Adler required a generalization

$$(\gamma^\mu D_\mu - m)\psi = (\gamma^\mu p_\mu + \zeta_\mu e_0 - i s_{\mu\nu}\phi^\nu - m)\psi = 0$$

of the Dirac equation to reduced to the Dirac equation:

$$\gamma^\mu(p_\mu - \phi_\mu)\psi = m\psi.$$

This yields $\zeta_\mu = -\phi_\mu/2$. However, the Dirac equation can be obtained with $\zeta_\mu = \phi_\mu/2$ also, from another generalization

$$(D_\mu \gamma^\mu - m)\psi = 0,$$

of the Dirac equation. Thus, such criteria are arbitrary and ad hoc, lacking a sound reasoning. The value $\zeta_\mu = \phi_\mu/2$ in the transportation rule resulting in (16) is based on a systematic reasoning as above.

## IV. Gauge group formulation of spinors

In this section, we express the transportation property stated in (16) in terms of a gauge group as defined in Sec. II. The indices $\mu,\nu,\lambda,\eta...$ refer to $\mathcal{M}_4$; $\bar{\mu},\bar{\nu},\bar{\lambda},\bar{\eta}....$, to the space part of $\mathcal{M}_4$, i.e., taking values $1,2,3$; $\alpha',\beta',\gamma',\delta'...$, refer to the gauge group; $\alpha,\beta,\gamma,\delta...$, refer to the spinor group manifold; which will be subdivided further into $\tilde{\alpha},\tilde{\beta},\tilde{\gamma},\tilde{\delta}...$ and $\hat{\alpha},\hat{\beta},\hat{\gamma},\hat{\delta}...$, defined in (18) below.

From (15), we have

$$\frac{1}{2}\left[\gamma_{\bar{\mu}},\gamma_{0}\right] \;=\; \delta_{\bar{\mu}}^{\tilde{\alpha}}\begin{bmatrix}\sigma_{\tilde{\alpha}} & 0 \\ 0 & -\sigma_{\tilde{\alpha}}\end{bmatrix} \;=\; \delta_{\bar{\mu}}^{\tilde{\alpha}}\,e_{\tilde{\alpha}};$$

$$\frac{1}{2}\left[\gamma_{\bar{\mu}},\gamma_{\bar{\nu}}\right] \;=\; -i\,\varepsilon_{\bar{\mu}\bar{\nu}}{}^{\hat{\alpha}}\begin{bmatrix}\sigma_{\hat{\alpha}} & 0 \\ 0 & \sigma_{\hat{\alpha}}\end{bmatrix} \;=\; -i\,\varepsilon_{\bar{\mu}\bar{\nu}}{}^{\hat{\alpha}}\,e_{\hat{\alpha}}; \qquad (18)$$

for $\tilde{\alpha},\hat{\alpha}=1,2,3$. Commutation properties of $\{e_\alpha\}$ can be obtained from those for the Pauli matrices defined in (6):

$$\begin{aligned}
\left[e_{\tilde{\alpha}},e_{\tilde{\beta}}\right] &= 2i\,\varepsilon_{\tilde{\alpha}\tilde{\beta}}{}^{\hat{\gamma}}\,e_{\hat{\gamma}} = C_{\tilde{\alpha}\tilde{\beta}}{}^{\hat{\gamma}}\,e_{\hat{\gamma}}, \\
\left[e_{\hat{\alpha}},e_{\hat{\beta}}\right] &= 2i\,\varepsilon_{\hat{\alpha}\hat{\beta}}{}^{\hat{\gamma}}\,e_{\hat{\gamma}} = C_{\hat{\alpha}\hat{\beta}}{}^{\hat{\gamma}}\,e_{\hat{\gamma}}, \\
\left[e_{\tilde{\alpha}},e_{\hat{\beta}}\right] &= 2i\,\varepsilon_{\tilde{\alpha}\hat{\beta}}{}^{\tilde{\gamma}}\,e_{\tilde{\gamma}} = C_{\tilde{\alpha}\hat{\beta}}{}^{\tilde{\gamma}}\,e_{\tilde{\gamma}}.
\end{aligned} \qquad (19)$$

Thus, $\{e_\alpha\}=\{e_{\tilde{\alpha}},e_{\hat{\alpha}}\}$ provides a basis for the 6-dimensional algebra generated by $\left[\gamma_\mu,\gamma_\nu\right]=-4is_{\mu\nu}$. The metric is usually taken to be a constant multiple of the Killing tensor

$$g'_{\alpha\beta} \;=\; C_{\alpha\delta}{}^{\gamma}\,C_{\gamma\beta}{}^{\delta} \;=\; 8\,g'_{\alpha\beta};$$

where $[g'_{\alpha\beta}]$ is the identity matrix of rank 6. Clearly, there is some freedom in selecting the metric. However, the results with all suitable metrics are essentially equivalent. We take the negative identity matrix for the metric $\left[g_{\alpha\beta}\right]$ on the spinor group manifold for consistency with other results. It is clear from (19) that $C_{\gamma\alpha\beta}=g_{\gamma\delta}C_{\alpha\beta}{}^{\delta}$ are completely antisymmetric with respect to its subscripts i.e., $C_{\gamma\alpha\beta}=-C_{\alpha\gamma\beta}=C_{\alpha\beta\gamma}$.

Now we have

$$\begin{aligned}
\frac{1}{2}dx^\mu\left[\gamma_\mu,\gamma_\nu\right]\phi^\nu &= \frac{1}{2}\left\{dx^0\left[\gamma_0,\gamma_{\bar{\nu}}\right]\phi^{\bar{\nu}}+dx^{\bar{\mu}}\left[\gamma_{\bar{\mu}},\gamma_0\right]\phi^0++dx^{\bar{\mu}}\left[\gamma_{\bar{\mu}},\gamma_{\bar{\nu}}\right]\phi^{\bar{\nu}}\right\} \\
&= -\;dx^0\,\varphi_0^{\tilde{\alpha}}\,e_{\tilde{\alpha}}+\;dx^{\bar{\mu}}\,\varphi_{\bar{\mu}}^{\tilde{\alpha}}\,e_{\tilde{\alpha}}+\;dx^{\bar{\mu}}\,\varphi_{\bar{\mu}}^{\hat{\alpha}}\,e_{\hat{\alpha}} \qquad (20)\\
&= dx^\mu\,\varphi_\mu^\alpha\,e_\alpha;
\end{aligned}$$

where

$$\varphi_0^{\tilde{\alpha}} = -\phi^{\bar{\mu}} \delta_{\bar{\mu}}^{\tilde{\alpha}}, \quad \varphi_{\bar{\mu}}^{\tilde{\alpha}} = \phi^0 \delta_{\bar{\mu}}^{\tilde{\alpha}}, \quad \varphi_{\bar{\mu}}^{\hat{\alpha}} = -i\phi^{\bar{\nu}} \varepsilon_{\bar{\mu}\bar{\nu}}{}^{\hat{\alpha}}.$$

From the arbitrariness of $dx^\mu$ in (20) or by direct computation, we have

$$\varphi_\mu^\alpha e_\alpha = \frac{1}{2}\left[\gamma_\mu, \gamma_\nu\right]\phi^\nu = -2i\, s_{\mu\nu}\phi^\nu \qquad (21)$$

It now follows from (16) and (20) that the change $\delta\psi$ in a 4-component spinor $\psi$ undergoing an infinitesimal parallel displacement in $\mathcal{M}_4$ is given by

$$\begin{aligned}\delta\psi &= \frac{1}{2}dx^\mu\left[\phi_\mu e_0 - 2i\, s_{\mu\nu}\phi^\nu\right]\psi \\ &= \frac{1}{2}dx^\mu\left[\phi_\mu e_0 + \varphi_\mu^\alpha e_\alpha\right]\psi = dx^\mu \varphi_\mu^{\alpha'} e_{\alpha'}\psi\end{aligned} \qquad ; \qquad (22)$$

where $\varphi_\mu^0 = \phi_\mu$. The identity element $e_0$ provides a basis in the multiplicative gauge group and thus, the augmented set $\{e_{\alpha'}\}_{\alpha'=0}^6 = \{e_0, e_\alpha\}_{\alpha=1}^6$ constitutes a basis in the composite gauge group. From (22), $\psi$ transforms as the "length" of a vector in the Weyl space obtained from $\mathcal{M}_4$ by coupling it with a multidimensional gauge group.

It is clear from (4) and (5) that the vectors in this Weyl space can also be treated in the same manner as the spinors. The basis in case of the vectors is provided by the basic Lorentz rotations [17]. Furthermore, it can be checked that the higher rank tensors also share this property with the spinors and vectors, and thus, the following methods are applicable and results extendible to them also. If a Weyl space is constructed starting with a general Riemannian space as the base, the affine connections are modified by adding the connections on the base given by the Christopher symbols. As is known, this creates an obstacle to expressing the displacement properties in terms of clearly defined Lie algebra elements. However, the following results are easily extendible to include a general Riemannian space as the base, as long as the structure group is a well defined gauge group.

## V.  Classical description of motion

The displacement property $\delta l = \phi_\mu dx^\mu l$ of the length of a vector expresses $\delta l$ in terms of the connection 1-form $\phi_\mu dx^\mu$ of a fiber bundle with $\mathcal{M}_4$ as the base and one dimensional gauge group as the vertical space. This bundle has been used as the underlying Riemannian space to describe a scalar particle classically and to obtain its quantum description by the path-integral method yielding the Klein-Gordon equation in an electromagnetic field [19]. Here we extend this approach to the spinors.

It is clear from (22), that the change $\delta\psi$ experienced by a Weyl spinor under an infinitesimal parallel displacement is expressed in terms of the connection forms of a fiber bundle with $\mathcal{M}_4$ as the base and the vertical space being the 7-dimensional composite gauge group manifold. The metric for this fiber bundle is given by (1). For a better transparency of the results, we label the coordinates used in Sec. II as

$$\left\{x^a\right\}_{a=4}^{10} = \left\{x^4, x^a\right\}_{a=5}^{10} = \left\{\chi^0, \chi^\alpha\right\}_{\alpha=1}^{6} = \left\{\chi^0, \chi^{\tilde{\alpha}}, \chi^{\hat{\alpha}}\right\}_{\tilde{\alpha},\hat{\alpha}=1}^{3}.$$

All of these coordinate labels will be used as convenient. Metric on the spinor group will be augmented to the negative identity matrix on the gauge group. In keeping with the requirement that the multiplicative group be a circle, the corresponding coordinate $\chi^0$ varies over a closed bounded interval in the real line. The other group coordinates vary over the spinor group. For convenience, we take the coordinate system on the spinor group manifold to be locally geodetic characterized by [4]

$$\partial_\gamma \varphi_\nu^\alpha + C_{\gamma\delta}{}^\alpha \varphi_\nu^\delta = 0; \qquad (23)$$

where $\partial_\gamma = \partial/\partial\chi^\gamma$. This choice of the coordinates simplifies some manipulations without compromising generality. The potentials $\varphi_\mu^\alpha$ depend on the group coordinates but $\varphi_\mu^0 = \phi_\mu$ is assumed to be independent of them, which is in fact included in (23) as the corresponding structure constants are all equal to zero. The arclength will be scaled for compatibility with the 1-dimensional case in absence of the spin, where the arclength $d\tau$ is defined by [19]

$$d\tau^2 = dx^\mu dx_\mu - (d\chi^0 + \phi_\mu dx^\mu)^2. \qquad (24)$$

Resulting arclength in the composite Kaluza-Klein manifold is given by

$$\begin{aligned} d\tau^2 &= g_{\mu\nu} dx^\mu dx^\nu - \left(d\chi^0 + \phi_\mu dx^\mu\right)^2 + g_{\alpha\beta}\left(d\chi^\alpha + \varphi_\mu^\alpha dx^\mu\right)\left(d\chi^\beta + \varphi_\nu^\beta dx^\nu\right) \\ &= dx^a dx_a \end{aligned} \qquad (25)$$

The results can be easily adjusted to accommodate a different scaling.

The equation of a geodesic can be determined by its variational characterization with $m\sqrt{\dot{x}^a \dot{x}_a}$ as the Lagrangian where $m$ is a constant and the dot denotes the derivative with respect to the parameter used to parameterize the trajectories, usually the arclength.

Being a homogeneous Lagrangian, this yields a parameter independent value of the action, which is quite suitable for a fixed trajectory but not for a collection of trajectories as is the case with the path-integrals. This can be corrected by treating the arclength $\tau$ as an independent parameter that can accommodate a collection of trajectories, and by replacing $m\sqrt{\dot{x}^a \dot{x}_a}$ by an equivalent inhomogeneous Lagrangian, often taken to be $m\dot{x}^a \dot{x}_a /2$ [14]. The action $S$ can be evaluated by solving the Hamilton-Jacobi equation, which now depends on $\tau$. A convenient way for the present to recover the action corresponding to the homogeneous Lagrangian is by setting $\partial S/\partial \tau = 0$. This yields the action equal to half of the value obtained by integrating $m\sqrt{\dot{x}^a \dot{x}_a}$ along the geodesic. To obtain the correct value, one can take $m(\dot{x}^a \dot{x}_a + 1)/2$ for the Lagrangian $L$ [15,19], which for the present case reads:

$$L = \frac{1}{2} m \left[ \dot{x}^\mu \dot{x}_\mu - \left(\dot{x}^0 + \phi_\mu \dot{x}^\mu\right)^2 + g_{\alpha\beta}\left(\dot{x}^\alpha + \varphi^\alpha_\mu \dot{x}^\mu\right)\left(\dot{x}^\beta + \varphi^\beta_\nu \dot{x}^\nu\right) + 1 \right]. \quad (26)$$

The momentums $\partial L/\partial \dot{x}^a$ are now given by

$$\begin{aligned}
q_0 &= -m\left(\dot{x}^0 + \phi_\mu \dot{x}^\mu\right), \quad q_\alpha = m\, g_{\alpha\beta}\left(\dot{x}^\beta + \varphi^\beta_\nu \dot{x}^\nu\right); \\
p_\mu &= m\left[\dot{x}_\mu - \phi_\mu\left(\dot{x}^0 + \phi_\nu \dot{x}^\nu\right) + g_{\alpha\beta}\varphi^\alpha_\mu\left(\dot{x}^\beta + \varphi^\beta_\nu \dot{x}^\nu\right)\right] \\
&= \left(m\dot{x}_\mu + \phi_\mu q_0 + \varphi^\alpha_\mu q_\alpha\right);
\end{aligned} \quad (27)$$

which are solved to yield the velocities $\dot{x}_a$:

$$\begin{aligned}
\dot{x}_\mu &= \frac{1}{m}\left(p_\mu - \phi_\mu q_0 - \varphi^\alpha_\mu q_\alpha\right); \\
\dot{x}_0 &= -\frac{1}{m}\left[q_0 + \phi^\mu\left(p_\mu - \phi_\mu q_0 - \varphi^\alpha_\mu q_\alpha\right)\right], \\
\dot{x}_\alpha &= \frac{1}{m}\left[q_\alpha - g_{\alpha\beta}\varphi^\beta_\nu g^{\nu\mu}\left(p_\mu - \phi_\mu q_0 - \varphi^\alpha_\mu q_\alpha\right)\right].
\end{aligned} \quad (28)$$

From (27) and (28) the Hamiltonian $H = \left(\dot{x}_a p^a - L\right)$ is given by

$$H = \frac{1}{2m}\left[g^{\mu\nu}\left(p_\mu - \phi_\mu q_0 - \varphi^\alpha_\mu q_\alpha\right)\left(p_\nu - \phi_\nu q_0 - \varphi^\beta_\nu q_\beta\right) + q_0\, q^0 + q_\alpha\, q^\alpha\right] - \frac{m}{2}; \quad (29)$$

where $q$ denotes the momentum on vertical space and $p$, on the base.

The classical motion is described by Hamilton's equations $\partial H/\partial x^a = -\dot{p}_a$, yielding from (29) together with (28) and (23):

$$\dot{p}_\mu = \left[\phi_{\nu,\mu}\, q_0 + \varphi^\beta_{\nu,\mu}\, q_\beta\right] \dot{x}^\nu;$$
$$\dot{q}_0 = 0, \qquad (30)$$
$$\dot{q}_\alpha = q_\beta\, \dot{x}^\nu \left(\partial_\alpha \varphi^\beta_\nu\right) = q_\beta\, \dot{x}^\nu\, C^\beta_{\ \gamma\alpha}\, \varphi^\gamma_\nu = 2i\, \varepsilon_{\gamma\alpha}^{\ \ \beta}\, \varphi^\gamma_\nu\, q_\beta\, \dot{x}^\nu.$$

with $X_{,a} = \partial_a X = \partial X/\partial x^a$. Conventional form of the equation of motion can be obtained from (28) and (30), resulting in

$$m\, \ddot{x}_\mu + q_0\, f_{\mu\nu}\, \dot{x}^\nu + q_\alpha\, \widehat{F}^\alpha_{\mu\nu}\, \dot{x}^\nu = 0; \qquad (31)$$

where

$$f_{\mu\nu} = \left(\phi_{\nu,\mu} - \phi_{\mu,\nu}\right); \quad \widehat{F}^\alpha_{\mu\nu} = \left(\varphi^\alpha_{\nu,\mu} - \varphi^\alpha_{\mu,\nu} + 2i\, \varepsilon_{\beta\gamma}^{\ \ \alpha}\, \varphi^\beta_\mu\, \varphi^\gamma_\nu\right).$$

The equation of motion (31) should be supplemented by the equation for $\dot{q}_{\alpha'}$ given by (30), which implies that $q_0$, $q^0$ are constants of motion, and together with the antisymmetry of $\varepsilon_{\alpha\beta\gamma}$, that $(\dot{q}_\alpha q^\alpha + q_\alpha \dot{q}^\alpha) = 0$, i.e., $q_\alpha q^\alpha$ is also a constant of motion. However, the charge vector $q$ rotates as described by the relation

$$\dot{q}_\alpha = 2i\, \varepsilon_{\gamma\alpha}^{\ \ \beta}\, \varphi^\gamma_\nu\, q_\beta\, \dot{x}^\nu$$

in (30). The equation of motion (31) compares with similar equation available in literature [4].

Spin of the charge vector is more transparent in the Cartesian coordinate system, as follows. Since the metrics with elements $g_{\tilde{\alpha}\tilde{\beta}}$, $g_{\hat{\alpha}\hat{\beta}}$ and $g_{\bar{\mu}\bar{\nu}}$ are all equal to the negative identity in 3-dimensional Euclidean space, the scalar product of two vectors $a = a^\alpha e_\alpha$ and $b = b^\alpha e_\alpha$ on the group manifold can be expressed as the product of the Cartesian vectors:

$$a \bullet b = a^\alpha e_\alpha \bullet e_\beta b^\beta = g_{\alpha\beta}\, a^\alpha\, b^\beta = g_{\tilde{\alpha}\tilde{\beta}}\, a^{\tilde{\alpha}}\, b^{\tilde{\beta}} + g_{\hat{\alpha}\hat{\beta}}\, a^{\hat{\alpha}}\, b^{\hat{\beta}} = -\left(\tilde{a}\cdot\tilde{b} + \hat{a}\cdot\hat{b}\right); \qquad (32)$$

where $\tilde{a}, \tilde{b}$ and $\hat{a}, \hat{b}$ are the Cartesian vectors in 3-dimensional space with components $a^{\tilde{\alpha}}, b^{\tilde{\beta}}$, and $a^{\hat{\alpha}}, b^{\hat{\beta}}$, respectively. A large dot denotes the scalar product on the group manifold and the smaller one denotes the usual Cartesian scalar product. In the matrix representations of $\{e_\alpha\}$, $g_{\alpha\beta} = -Tr(e_\alpha e_\beta)/4$, and thus

$$a^\alpha \left(e_\alpha \bullet e_\beta\right) b^\beta = -\frac{1}{4} Tr\left(a^\alpha e_\alpha e_\beta b^\beta\right).$$

Similarly, $\varepsilon_{ijk} b^j c^k = (b \times c)_i$, yielding the identities

$$\varepsilon_{ijk} a^i b^j c^k = a \cdot b \times c; \quad \varepsilon_{ijk} \varepsilon_{\bar{\mu}\bar{\nu}}{}^k b^{\bar{\mu}} c^{\bar{\nu}} a^j = \varepsilon_{ijk} a^j (b \times c)^k = \left[a \times (b \times c)\right]_i; \quad (33)$$

where $a, b, c$ are the Cartesian vectors with components $a^i, b^j, c^k$, and cross denotes the vector product of 3-vectors.

From (30), (32) and (33) together with (27), we have

$$\begin{aligned}
\dot{q}_{\tilde{\alpha}} &= q_\beta \, \dot{x}^\nu \, C^\beta{}_{\gamma\tilde{\alpha}} \, \varphi^\gamma_\nu = 2\left[\tilde{q} \times (\dot{\bar{x}} \times \bar{\phi})\right]_{\tilde{\alpha}} + 2i\left[\hat{q} \times (\phi^0 \dot{\bar{x}} - \dot{x}^0 \bar{\phi})\right]_{\tilde{\alpha}} \\
\dot{q}_{\hat{\alpha}} &= q_\beta \, \dot{x}^\nu \, C^\beta{}_{\gamma\hat{\alpha}} \, \varphi^\gamma_\nu = 2\left[\hat{q} \times (\dot{\bar{x}} \times \bar{\phi})\right]_{\hat{\alpha}} + 2i\left[\tilde{q} \times (\phi^0 \dot{\bar{x}} - \dot{x}^0 \bar{\phi})\right]_{\hat{\alpha}}
\end{aligned} \quad (34)$$

where $\tilde{q}, \hat{q}$ are the vectors with components $q^{\tilde{\alpha}}, q^{\hat{\alpha}}$, respectively; and $\bar{\phi}, \bar{x}$ are the vectors with components $\phi^{\bar{\mu}}, x^{\bar{\mu}}$, respectively. From (34), $q$ is pictured as a classical spinning charge. The term $\left[\hat{q} \times (\dot{\bar{x}} \times \bar{\phi})\right]$ corresponds to the spin in space. The spinning charge vector, $\hat{q}$ can be expressed as $\hat{q} = e\hat{u}$, where $e$ is the magnitude of charge and $\hat{u}$ is a rotating unit vector in 3-dimensional Euclidean space. The vector $\tilde{q}$ can also be expressed as $\tilde{q} = e\tilde{u}$, with a similar meaning. Equations similar to (31) and (34) were conjectured by Wong based on a quantum field theoretical formulation of the isotopic spin [20].

Spin can also be expressed in the compact matrix form by defining $Q = q^\alpha e_\alpha$ and using (21), yielding,

$$\begin{aligned}
\dot{Q} &= \dot{q}^\alpha e_\alpha = q^\beta \, \dot{x}^\nu \, C_{\beta\gamma}{}^\alpha e_\alpha \, \varphi^\gamma_\nu = q^\beta \, \dot{x}^\nu \left[e_\beta, e_\gamma\right] \varphi^\gamma_\nu \\
&= \dot{x}^\nu \left[Q, e_\gamma\right] \varphi^\gamma_\nu = \frac{1}{2} \dot{x}^\mu \left[Q, \left[\gamma_\mu, \gamma_\nu\right]\right] \phi^\nu \\
&= 2i \, \dot{x}^\mu \left[s_{\mu\nu}, Q\right] \phi^\nu.
\end{aligned} \quad (35)$$

There is not much to be gained by expressing the equation of motion (31) in the Cartesian system; instead we express it in its compact matrix form. We have

$$q_\alpha \hat{F}^\alpha_{\mu\nu} = q_\alpha \left( \varphi^\alpha_{\nu,\mu} - \varphi^\alpha_{\mu,\nu} + 2i\, \varepsilon_{\beta\gamma}{}^\alpha \varphi^\beta_\mu \varphi^\gamma_\nu \right)$$

$$= 2i\, Q \bullet \left( s_{\mu\lambda}\, \phi^\lambda{}_{,\nu} - s_{\nu\lambda}\, \phi^\lambda{}_{,\mu} + 2i \left[ s_{\mu\lambda}, s_{\nu\eta} \right] \phi^\lambda \phi^\eta \right).$$

Substituting this in (31) results in

$$m\, \ddot{x}_\mu + q_0\, f_{\mu\nu}\, \dot{x}^\nu + 2iQ \bullet \left( s_{\mu\lambda}\phi^\lambda{}_{,\nu} - s_{\nu\lambda}\phi^\lambda{}_{,\mu} + 2i \left[ s_{\mu\lambda}, s_{\nu\eta} \right] \phi^\lambda \phi^\eta \right) \dot{x}^\nu = 0. \quad (36)$$

## VI.  Quantum description of spinors

In the path-integral formulation, the wavefunction $\psi$ is obtained from the following integral equation [6,7]:

$$\psi(x, \tau + \varepsilon) = \int dy\, v(y)\, \exp(i\, S[x, \tau + \varepsilon; y, \tau])\, \psi(y, \tau), \quad (37)$$

where $S[x, \tau + \varepsilon; y, \tau]$ is the classical action along an extremal from a point $(y, \tau)$ on the trajectory to another one $(x, \tau + \varepsilon)$, and $v(y)$ is a function depending on the underlying geometry. For the curved spaces such as the present Weyl space, $v(y) = v'\sqrt{|\hat{g}(y)|}$, where $|\hat{g}(y)|$ is the determinant of metric $\hat{g}(y)$ with $v'$ being independent of $y$. The quantum mechanical equation for $\psi$, is obtained by expanding both sides of (37) in powers of $\sqrt{\varepsilon}$ and comparing the first few terms, yielding a Schrödinger type equation. A convenient expression was obtained by Cheng with the Lagrangian taken to be $m\dot{x}^a \dot{x}_a / 2$ [14], which can be adjusted for the present Lagrangian $m(\dot{x}^a \dot{x}_a + 1)/2$ to read

$$-2mi\, \frac{\partial \psi}{\partial \tau} = \left[ \frac{1}{\sqrt{|\hat{g}|}}\, \partial_a \left( \sqrt{|\hat{g}|}\, \hat{g}^{ab}\, \partial_b \right) + m^2 - \frac{1}{3} R \right] \psi; \quad (38)$$

where $R$ is the curvature scalar. Derivations of alternative expressions of (38) are available elsewhere [7,13,15].

As pointed out above, the required value of the principal function can be recovered by setting $\partial S / \partial \varepsilon = 0$. It follows by differentiating (37) with respect to $\varepsilon$, and setting $\partial S / \partial \varepsilon = 0$, that $\partial \psi / \partial \tau = 0$. With this (38) reduces to the Klein-Gordon equation extended to a general Reimannian space:

$$\left[ \frac{1}{\sqrt{|\hat{g}|}} \partial_a \left( \sqrt{|\hat{g}|}\, \hat{g}^{ab}\, \partial_b \right) + m^2 - \frac{1}{3} R \right] \psi = 0; \tag{39}$$

Substituting for $\hat{g}$ from (1), we obtain

$$g^{\mu\nu}\left(\partial_\mu - \phi_\mu d_0 - \varphi_\mu{}^\alpha \partial_\alpha\right)\left(\partial_\nu - \phi_\nu d_0 - \varphi_\nu{}^\beta \partial_\beta\right)\psi + d_0 d^0 \psi + \partial_\alpha \partial^\alpha \psi + \left(m^2 - R'\right)\psi = 0; \tag{40}$$

where $d_0$ denotes the derivatives with respect to $x^0$; $R' = R/3$ with

$$R = \frac{1}{4}\left(f_{\mu\nu} f^{\mu\nu} + \hat{F}^\alpha_{\mu\nu} \hat{F}^{\mu\nu}_\alpha\right);$$

and $f_{\mu\nu}$, $\hat{F}^{\bar{\alpha}}_{\mu\nu}$ are as defined in (31). Derivation of (40) can also be based on the alternative expressions of (39).

From the Peter-Weyl theorem, characters of the irreducible representations of a group constitute a complete orthogonal system providing a basis to expand a function on the group, e.g., for the multiplicative gauge group, which is a circle, the characters and hence a basis, is given by the set $\left\{\exp\left[-ienx^0\right]\right\}$, thereby providing the Fourier series expansion of $\psi$ [18]. This eliminates $d_0$ from (40) and yields the charge quantization. We consider the term corresponding to $n=1$. The function $\psi$ can be expanded further in terms of the characters of spinor group, which can be expressed in terms of SU(2). However, such a treatment is somewhat diversionary. Instead, we reduce (40) by using the fact that on the group the derivatives are given by its Lie algebra elements:

$$\partial_\alpha \, Exp\left(-i e\, x^\beta e_\beta\right) = -i e\, e_\alpha\, Exp\left(-i e\, x^\beta e_\beta\right),$$

where $Exp$ denotes the generalized exponential, which can be obtained by multiplying the elements close to the identity. A spinor $\psi$ under a group operation transforms from $\psi$ to $Exp\left(-iex^\beta e_\beta\right)\psi$, thus yielding $\partial_\alpha \psi = -ie\, e_\alpha \psi$. This corresponds to the basic coefficient in an expansion. Substitutions in (40) yields

$$\begin{aligned} g^{\mu\nu}\left(\partial_\mu + ie\, \phi_\mu\right)\left(\partial_\nu + ie\, \phi_\nu\right)\psi + e\, s^{\mu\nu} f_{\mu\nu} \psi \\ + 4e\, s^{\mu\nu}\phi_\nu \partial_\mu \psi + 4e^2 s^{\mu\nu} s_{\mu\lambda}\phi_\nu \phi^\lambda \, \psi + \left(m^2 + 7e^2 - R'\right)\psi = 0. \end{aligned} \tag{41}$$

The term $es^{\mu\nu} f_{\mu\nu}$ in (41) having no classical analogue appears also in the squared Dirac equation. The corresponding equation (41) for the Weyl spinor, has the following additional terms: The term $\left(4e\, s^{\mu\nu}\phi_{,\nu}\partial_{\mu}\psi\right)$ corresponds to the classical spinning charge described in (34) and (35). The term $\left(4e^2 s^{\mu\nu} s_{\mu\lambda}\phi_{,\nu}\phi^{\lambda}\right)$ is manifest in the classical Hamiltonian given by (29) and appears in the definition of field in the classical description of motion provided by (31) and (36). The curvature scalar $R'$ is a characteristic of the path-integration in curved spaces and $\left(7e^2\right)$ is the result of the Laplacian on the structure group appearing in (40), which is a characteristic of the Kaluza-Klein formulation. Without these additional terms, (41) reduces to the squared Dirac equation.

## VII. Concluding remarks

This paper presents a study of the 4-component spinors of first rank that result from the vectors in the Weyl space constructed from the Minkowski space by abandoning the requirement of length invariance under parallel transport. An operation of parallel displacement on the vectors in this affinely connected space is shown to be the product of a multiplicative gauge group and an orthochronous, proper Lorentz transformation, which admits a spinor representation thereby giving rise to the spinors. Displacement properties of the spinors induced by the vectors are shown to be equivalent to the coupling of a particle with a multidimensional gauge field, which is formulated in the Kaluza-Klein framework with fiber bundle construction equipped with a Riemannian structure. This enables the classical description of spinors as particles in a curved Riemannian space. In its classical description, a Weyl spinor moves as a spinning charged particle. Quantum description of the spinors is then obtained by the method of path-integration in curved spaces. The resulting Klein-Gordon equation for the Weyl spinor that incorporates explicitly its classical features in its quantum version differs from the squared Dirac equation by a few terms. This treatment of spinors is extendible to the vectors and tensors of higher rank. The procedure used here to obtain the descriptions of motion is also applicable to treat the particles in Riemannian spaces coupled to multidimensional gauge fields.